\documentstyle[12pt]{article}
\def\graphic #1#2#3#4#5{

    \noindent
    \hrulefill
    \vskip#2 \relax
    \vskip -3.9 cm
    \hskip 4.8 cm
    {\large \bf Universidade do Estado do Rio de Janeiro }
    \newline

    \vskip -0.25 cm
    \hskip 7.5 cm
    {\large \bf Instituto de F{\'\i}sica }

    \vskip 1 cm
    \hskip 7.5 cm
    {\large Phys-Pub #4 }

    \hskip 7.5 cm
    {\large Preprint}

    \hskip 7.5 cm
    {\large #5 }

    \medskip
    \noindent
    \hrulefill

    \vskip 2.9 cm
    }
\def\({\c c}
\def\|{\'\i}

\def\eq #1{Eq. (\ref{#1})}
\def\l{\left}
\def\r{\right}
\def\pa{\partial}
\def\fr{\frac}
\def\|{\'\i}
\def\*{\&\!*}
\def\p*{\&p}
\def\C{{\it commutative}}
\def\NC{{\it non-commutative}}

\def\m{{\vrule height0.65em width0em depth0.65em}}

\begin{document}
\hspace\parindent
\thispagestyle{empty}

\graphic{2 in}{1.6 in}{uerj.bmp}{027/95}{October 1995}
\centerline{\LARGE \bf Symbolic Computing with}

\medskip
\centerline{\LARGE \bf  Grassman Variables}

\bigskip
\bigskip
% Authors:
\centerline{\large
E.S. Cheb-Terrab\footnote{Departamento de F\'{\i}sica Te\'orica, IF-UERJ.
E-mail: terrab@vmesa.uerj.br.}
%?\footnotemark[1]
%and ?\footnote{?}
}

\bigskip
\bigskip
\bigskip
\begin{abstract}

A package of Maple 5.3 commands for doing calculations with
anticommutative variables is presented.

\end{abstract}

\bigskip
\centerline{ \underline{\hspace{6.5 cm}} }

\medskip
% \centerline{ {\bf To be published in the Proceedings of }}
%
% \centerline{ {\bf {\it Computing in High Energy Physics, CHEP-1995}}}
%
% \centerline{{\bf World Scientific Publishing Co.}}
\centerline{ {\bf (Submitted to Journal of Symbolic Computation)} }

\newpage
\section{Introduction}

\noindent Working with anticommutative symbolic objects is becoming part
of the standard activity in mathematical-physics (Sohnius, 1985). On the
other hand, almost all available general-purpose symbolic computing
systems have not yet included anticommutative variables (ACVs) in their
default computational domains. In Maple, for instance, this is reflected
in the product operator, $`*`$, defined as commutative, and in the
standard differentiation command, {\bf diff}, not designed to evaluate
derivatives with w.r.t ACVs. Also, the admitted non-commutative product
operator, $\*$, does not have its differentiation rule and its basic
simplification operations defined.

This paper presents a set of Maple 5.3 commands for working with ACVs,
covering the topics mentioned above. The exposition is organized as
follows. In Sec.\ref{package}, a brief review of the package is presented.
In Sec.\ref{acproduct}, the  extension of the computational domain and the
rules for manipulating and simplifying anticommutative products are
discussed. Sec.\ref{gdiff} is dedicated to the differentiation problem;
that is, the rules for differentiating anticommutative products, composite
functions, and for obtaining high order derivatives w.r.t. ACVs.
Sec.\ref{examples} contains a brief illustration of how the new commands
work. Finally, the Conclusions contain some general remarks about this
work and its possible extensions.

\section{Anti-commutative variables and the {\it Grassman} \newline package}
\label{package}

To start with, let's recall that any pair $\{\theta_i,\,\theta_j\}$,
belonging to a set $\{\theta_1,...,\theta_n\}$ of {\it anticommutative}
variables, satisfies\footnote{Following Maple conventions, the symbol used
by the package's commands to represent a \NC\  product operator is $\*$.}
$\theta_i \* \theta_j = -\theta_j \* \theta_i$. As a convention, the {\bf
parity} is here defined as $1$ for all $\theta_n$ variables, and $0$ for
any other ({\it commutative}) variable. These definitions are extended to
include the concepts of {\it commutative}, $F(...)$, and {\it
anticommutative}, $Q(...)$, functions, with  parity defined as $0$ and
$1$, respectively. Note that both $F(...)$ and $Q(...)$ {\it may or not}
depend on the anticommutative variables $\theta_n$.

The package of commands here presented allows its user to work with
{\it anticommutative} variables and functions by introducing algebraic
symbols to represent them, and by defining their basic operations and
differentiation rules taking into account their {\it anticommutative}
character. A brief review of the package is as follows:

\begin{description}

\item{-} all variables built with the string {\tt `theta`} followed by a
positive number are considered {\it anticommutative} variables; all
functions with name beginning with the capital letter {\tt Q} are
considered {\it anticommutative} functions;

\item{-} three type subroutines, {\bf `type/grassman`}, {\bf
`type/commutative`} and {\bf `type/anti\-commutative`}, and a command,
{\bf parity}, permit the recognition of the {\it commu\-tative} and {\it
anticommutative} character of any algebraic symbol or expression;

\item{-} the {\bf midentity} command replaces nested, not simplified, {\it
non-commutative} $\*$ products, in a given expression, by ``un-nested" and
simplified ones;

\item{-} the {\bf msort} command sorts the operands of each $\*$ product
of a given expression in a unique manner, permitting an efficient
manipulation of the expression by the Maple system as a whole;

\item{-} the {\bf mexpand} command sorts (using {\bf msort}) and
recursively expands the $\*$ products over sums in a given expression;

\item{-} the {\bf gdiff} command is a differentiation command designed to
work both with commutative and anticommutative variables.

\item{-} the set of commands {\bf useD}, {\bf usegdiff} and {\bf usediff}
change the format of the derivatives of a given expression to the {\bf D},
{\bf gdiff} and {\bf diff} format, respectively;

\item{-} the utility subroutines {\bf `diff/\&*`} and {\bf
`diff/function`} define the rules for differentiating $\*$ products and
composite functions, taking into account the possible {\it
anticommutative} character of the symbols involved.

\end{description} Of the eight commands of the package, {\bf mexpand} and
{\bf gdiff} are the key ones. The former makes calls to {\bf msort}, which
in turn makes calls to {\bf midentity}, producing a somewhat elaborated
result, while {\bf gdiff} is a generalization of the standard {\bf diff}
command; working as the latter concerning syntax and display and when the
differentiation variable is not an ACV.

\section{The computational domain and the {\it non-commuta\-tive} $\*$
products} \label{acproduct}

\subsection{The {\bf `type/...`} subroutines}

As mentioned in the Introduction, the default computational domain of the
Maple system does not include anticommutative variables. Hence, the
departure point was the development of three subroutines, {\bf
`type/grassman`}, {\bf `type/commutative`}, {\bf `type/an-ticommutative`},
and a command, {\bf parity}, to recognize the possible
commutative/anti\-com\-mutative character of a given expression. The {\bf
`type/grassman`} subroutine identifies an anticommutative variable or
function, as follows\footnote{In what follows, the {\it input} can be
recognized by the Maple prompt \verb->-.}:

\medskip
\noindent {\it Calling sequence:}
\begin{verbatim}
> type(var,grassman);
\end{verbatim}
\noindent {\it Scheme for the answers:}
\begin{description}

\item{-} {\it if} {\tt var} is a string {\it and} the first five characters
are {\tt theta} {\it and} the remaining part of the string is a positive
integer {\it then} {\bf true}

\item{-} {\it if} ( {\tt var} is a string {\it and} the first character is
{\tt Q} ) {\it or} ( ( {\tt var} is a function {\it or} an indexed object
) {\it and} the first character of its name\footnote{In Maple, the name of
such objects is given by {\tt > op(0,var);}} is {\tt Q} ) {\it then} {\bf
true}

\item{-} {\it else} {\bf false}

\end{description}

The {\bf `type/commutative`} subroutine addresses a more general task,
which is to determine whether a whole algebraic expression, arbitrarily
composed with sums, products, powers, functions and constants, {\it is or
is not} commutative; it relies completely on the {\bf `type/grassman`}
subroutine, and works as follows:

\medskip
\noindent {\it Calling sequence:}
\begin{verbatim}
> type(expr,commutative);
\end{verbatim}
\noindent {\it Scheme for the answers:}
\begin{description}

\item{-} {\it if} {\tt expr} is a sum {\it then} the subroutine makes a
recursive calling of itself {\it and if} each operand is commutative {\it
then} {\bf true}

\item{-} {\it if} {\tt expr} is a product {\it then} the subroutine makes a
recursive calling of itself {\it and if} there is an even number of
non-commutative operands {\it then} {\bf true}

\item{-} {\it if} {\tt expr} is a power {\it then} {\it if} the base is
commutative or the exponent is even the {\bf true}

\item{-} {\it else} send the task to {\bf `type/grassman`}, returning {\bf
true} or {\bf false} correspondingly.

\end{description} Note that an expression is being recognized as {\it
commutative} just in the ``positive" case, that is, when this
commutativity is actually verified. For instance, the  {\bf
`type/anticom\-mu\-tative`} subroutine works in almost the opposite manner,
but an expression can be neither {\it commutative} nor {\it
anticommutative}; for example, a sum of objects of each type.

Finally, the {\bf parity} command is more like a ``macro", and works as
follows:

\medskip
\noindent {\it Calling sequence:}
\begin{verbatim}
> parity(expr);
\end{verbatim}
\noindent {\it Scheme for the answers:}
\begin{description}

\item{-} {\it if} {\tt expr} is of commutative type  {\it then} $0$

\item{-} {\it if} {\tt expr} is of anticommutative type {\it then} $1$

\item{-} {\it else} return the string {\it `undefined`}.

\end{description}

The three {\bf `type/...`} subroutines and the {\bf parity} command, are
recursively used by the other commands of the package before the decision
of what to do with the received arguments is taken. In this manner, though
it is not possible to introduce changes in Maple's kernel, it is possible
to introduce the tools for making correct evaluations of the
commutative/anticommutative character of an algebraic symbol. Then, the
correct results can be obtained by using the appropriate commands, mainly
$\*$ for the product of objects containing ACVs, {\bf mexpand} for
simplifying these products and {\bf gdiff} for evaluating derivatives.
% w.r.t ACVs (see below).

\subsection{The {\it non-commutative} $\*$}

To represent non-commutative products, the Maple system uses the $\*$
operator; it has the correct precedence with regard to the other basic
operations, but no elementary simplification or differentiation rules for
it are defined. Therefore, the idea was to take advantage of the
precedence rules of $\*$ and make it work as a {\it commutative} or {\it
anticommutative} product, according to the parity of its operands.

This is accomplished by the package's commands {\bf midentity}, {\bf
msort}, {\bf mexpand} and {\bf gdiff}, when they are called, by
substituting all occurrences of $\*$ by a subroutine $\p*$, which actually
realizes all the calculations. The idea was not to assign anything to the
$\*$ operator, in order to maintain the compatibility with the other
commands of the Maple Standard Library (MSL) which make calls to it.

Additionally, many usually desired simplifications of this $\*$ product
are automatically accomplished by the $\p*$ routine. These simplifications
are at the base of all the subsequent calculations, and can be summarized
as follows:

\medskip \noindent {\it Calling sequence:}
\begin{verbatim}
> &p(a,b,c,...);
\end{verbatim}

\noindent {\it Scheme for the answers:}
\begin{description}

\item{-} {\it if} there is only one operand, say {\tt a},  {\it then} {\tt a}

\item{-} {\it if} $0$ belongs to the sequence of operands {\it then} $0$

\item{-} {\it if} an operand, say {\tt a}, is in turn a product such as
$\*(f,g,h)$ or $fgh$, {\it then} replace it by its operands $ f,g,h$ and
{\it evaluate the whole expression again};\label{recurse}

\item{-} {\it else} return an expression of the form $\alpha\,\*\!(...)$,
were $\alpha$ is build as a standard product of all the operands of {\tt
\&p(a,b,c,...)} satisfying the condition of {\it not having grassman-type
variables or functions}.

\end{description} Note that the third item is applied recursively and,
as a consequence, an arbitrarily nested structure containing combinations
of $\*$ and $`*`$ will always be reduced to a plain structure, with all
the operands inside the $\*$ product on the same level, and with all the
operands not having grassman-type objects as factors outside. For example,
the process above will map nested products as in

$$
\*(a\,\*(F(\theta),b\,\*(Q_1(\theta),c\,Q_2(\theta))))
\rightarrow a\,b\,c\ \*(F(\theta),Q_1(\theta),Q_2(\theta))
$$

\noindent The $\p*$ operator may keep objects of parity $= 0$ inside a
$\*$ product, even when this objects actually commute with all the other
ones, for instance $F(\theta)$ in the example above. The reason for this
is related to the differentiation w.r.t $\theta_n$ variables (objects of
parity one), and can be seen by considering the product of a parity $=0$
function, say $F(\theta)$, times a parity $=1$ function, say $Q(\theta)$.
This product satisfies:

$$F(\theta)\ Q(\theta)=Q(\theta)\ F(\theta)$$ Differentiating the
left-hand-side (lhs), for example, one receives

$$
\fr{\pa F(\theta)}{\pa \theta}\ Q(\theta)
+ F(\theta)\ \fr{\pa Q(\theta)}{\pa \theta}
% =
% \fr{\pa Q(\theta)}{\pa \theta}\ F(\theta)
% - Q(\theta)\ \fr{\pa dF(\theta)}{\pa \theta}
$$

\noindent Since the operands of $\fr{\pa F(\theta)}{\pa \theta}\
Q(\theta)$ does not commute (both now have parity $=1$), this result is
correct {\it if one maintains the order of the operands}, that is, {\it if}
the product is represented using the \NC\  operator $\*$. On the other
hand, if one begins using the standard $`*`$ for representing the \C\
product $F\, Q$, then one should expect the result to be expressed in
terms of the \C\ $`*`$ too, which in turn will be wrong.

Thus, in the context of the package here presented, to obtain the correct
results, the $\*$ operator should be used to represent both commutative
and non-commutative products, while $\p*$ will take care of keeping as
arguments just the operands having grassman-type objects inside. The
surface command for achieving these simplifications is {\bf midentity}.

\subsubsection*{A ``canonical form" for $\*(a,b,c...)$}

In order to make more efficient the simplification of expressions
containing $\*$ products by the commands of the MSL, a ``canonical form"
for these products was implemented; that is, a ``unique" manner of writing
equivalent but apparently different such products. The command
accomplishing this process is {\bf msort}, which makes calls to {\bf
midentity} and works as follows:

\medskip
\noindent {\it Calling sequence:}
\begin{verbatim}
> msort(expr);
\end{verbatim}
\noindent {\it Scheme for the answer:}
All $\*$ products of {\tt expr} are sorted according to:

\begin{description}

\item{-} the sequence of arguments of $\*$ is split in two sequences, $s_1$
and $s_2$, containing all the \C\  objects, and all the \NC\ ones,
respectively;

\item{-} the elements of $s_1$ are sorted using machine ordering ({\tt
table(symmetric)})

\item{-} the elements of $s_2$ are sorted using machine ordering ({\tt
table(antisymmetric)}), and a $-1$ factor is introduced when the number of
required permutations is odd.

\end{description} Also, this mechanism automatically evaluates to $0$ all
$\*$ products containing powers of non-commutative symbols.

The {\bf mexpand} command addresses the problem of distributing $\*$
products over sums, makes calls to {\bf msort}, and also works
recursively. This distribution of $\*$ products over sums will be relevant
both in the process of sorting the $\*$ products explained above and in
taking advantage of the set of simplification commands of the MSL.

\section{Differentiation of expressions containing {\it grassman-type}
symbols}
\label{gdiff}

The introduction of anticommutative objects in the computational domain
required both the extension of the differentiation rules and the
definition of equivalencies between the standard {\bf D} and a new {\bf
gdiff} differentiation operators.

\subsection{Differentiation of $\*$ products, composite functions and high
order derivatives}

The differentiation operation was extended by defining the differentiation
rules for $\*$ products and composite functions, and high order
derivatives. This was accomplished by building a {\bf `diff/\&*`}
subroutine, adapting the standard {\bf `diff/function`}, and creating a
new differentiation command, {\bf gdiff}, which in turn relies completely
on the standard {\bf diff}.

The differentiation rule for products, {\bf `diff/\&*`}, was programmed to
return a result according to
$$
\fr {\partial} {\partial \theta} (A\, \* B) =
\fr {\partial A} {\partial \theta}\, \* B
+ (-1)^{\mbox{\footnotesize parity(A)}}\, A\, \* \fr {\partial B}
{\partial \theta}
$$

The differentiation rule for composite functions was programmed
introducing a few lines in the standard {\bf `diff/function`}, taking
care {\it not to change anything} in its standard behavior with
regard to commutative objects, but returning results according to:
$$
\fr {\partial} {\partial \theta}
\l(\m A\,(\, B\,)\,\r) =
\fr {\partial B} {\partial \theta}\ \* \fr {\partial A} {\partial B}
$$
when grassman-type objects are involved in the ``derivand".

To request the evaluation of a derivative taking into account the \NC\
character of the involved symbols one should use the {\bf gdiff} command.
The syntax of {\bf gdiff} is that of the standard {\bf diff}, and a {\bf
`print/gdiff`} procedure was created in order to obtain the same display
too. A brief summary of how this command works is as follows.

\medskip
\noindent {\it Calling sequence:}
\begin{verbatim}
> gdiff(expr,x);
\end{verbatim}

\noindent {\it Scheme for the answers:}
\begin{description}

\item{-} {\it if} ({\tt x} is not of grassman-type) {\it or} (there are no
derivatives within {\tt expr}) {\it then} send the task to the standard
{\bf diff} \label{casediff}

\item{-} {\it if} {\tt expr} is a derivative {\it then}
\begin{enumerate}

    \item split the sequence of differentiation variables into two
sequences, $s_{c}$ and $s_{nc}$, containing the commutative and
non-commutative variables, respectively;

    \item differentiate the derivand w.r.t the commutative variables $s_c$
using {\bf diff};

    \item sort the $s_{nc}$ elements using machine ordering, differentiate
the result of the previous step w.r.t the sorted $s_{nc}$ grassman
variables, using {\bf gdiff}, and introduce a $-1$ factor when an odd
number of permutations was required\footnote{To avoid infinite recursion,
the result of this step is returned ``built", but ``unevaluated".}.

\end{enumerate}

\item{-} {\it if} {\tt expr} contains, say $n$ derivatives {\it then}
\begin{enumerate}

    \item substitute the $n^{th}$ derivative by a function $Y_n(x)$,
where $Y$ is a local variable;

    \item differentiate the resulting expression w.r.t $x$, using {\bf
diff}, obtaining an expression involving the symbols $\fr{\pa}{\pa x}
Y_n(x)$;\label{derivate}

    \item differentiate the $n^{th}$ derivative found in {\tt expr} w.r.t
$x$, using {\bf gdiff}, and build a set of equivalencies $\{\fr{\pa}{\pa
x} Y_n(x)=R_n\}$, where $R_n$ represents the obtained results;

    \item introduce these equivalencies in the result of step
\ref{derivate}, sort all $\*$ products using {\bf msort} and return a
result

\end{enumerate}

\end{description}

\subsection{Equivalencies between the {\bf D} and {\bf gdiff}
differentiation operators}
\label{usegdiff}

The Maple system includes two differentiation formats for representing
derivatives, related to the {\bf diff} and the {\bf D} operators,
respectively. Although the first one is more intuitive, the second one is
more general: {\it all} derivatives can be represented using the {\bf D}
operator while {\it just some} using {\bf diff} (Monagan and Devitt,
1992). The conversion between formats in a given algebraic expression can
be realized by means of the {\bf convert} command.

Having the option of representing {\it all} derivatives in a unique
manner, as is the case of the {\bf D} format, is a highly desirable
property of a symbolic computing system. Hence, the idea was to define a
correspondence between the new {\bf gdiff} and the {\bf D} formats too.
With this purpose in mind, consider, for instance, the standard
representations for a second order derivative:

\begin{equation}
\fr{\pa^2}{\pa \theta_1\, \pa \theta_2}\,f(\theta_1,\theta_2,\theta_3)
= D_{1,2}(f)(\theta_1,\theta_2,\theta_3)
\label{equivalence}
\end{equation}

\noindent Since high order derivatives are supposed to commute, the order
in which $\theta_1$ and $\theta_2$ appear in the lhs, related to the {\bf
diff} format, is irrelevant and ``session dependent" (machine ordering).
On the other hand, the {\bf D} representation in the rhs carries an
``additional information": the ordering of the numbers $1$ and $2$, which
will always be numerical order. This property of the {\bf D} format can be
used to address the fact that, when differentiating w.r.t ACVs, the order
in which the derivatives are realized must be taken into account.
Therefore, the equivalence between the {\bf gdiff} and {\bf D} formats was
proposed as it appears in \eq{equivalence}, now admitting that the lhs was
built using {\bf gdiff}, thus preserving both the order in which
$\theta_1$ and $\theta_2$ are displayed and that in which the derivatives
are realized.

The conversion between equivalent formats was implemented by adapting the
{\bf useD} and {\bf usediff} commands of the {\it partials} package
(Cheb-Terrab, 1994) and building a new one, {\bf usegdiff}. The ``names"
these commands have resemble the ``actions" they realize, and the {\bf
`convert/...`} facility was not used in order to keep as much as possible
the compatibility with the other commands of the MSL\footnote{The
subroutines {\bf `convert/D`} and {\bf `convert/diff`} already exist and
work a bit differently from {\bf useD} and {\bf usediff}.}. Note also
that, while it was possible to state conversion rules between the {\bf
gdiff} and {\bf D} formats, this conversion is {\it not possible} between
{\bf gdiff} and {\bf diff}, since high order derivatives represented using
the latter are assumed by the whole system to commute.

\section{Examples}
\label{examples}
This section aims at giving a brief illustration of how the new commands
work. To start with, let's load the code and introduce some macros and
alias that improve the readability of the {\it input/output}\footnote{In
Maple, " means the last expression and "" means the second last
expression. Also, special care was taken to keep the {\it input}  and {\it
output} shown along this paper with almost exactly the same format and
aspect as that which appears on the computer screen\label{winmaple}.} of
$\theta_1$, $\theta_2$, the functions $Q_{1}(x,y,\theta 1,\theta 2)$,
$f_{1}(x,y,\theta 1,\theta 2)$, $f_{2}(x,y,\theta 1,\theta 2)$ and the
composite function $Q_{2}(\theta 1, \theta 2,Q_{1})$
\begin{verbatim}
> read grassman;
> t1=theta1, t2=theta2;
> map(macro,["]):
\end{verbatim}
$$
{\it t1}={ \theta 1}, {\it t2}={ \theta 2}
$$
\begin{verbatim}
> (Q[1]=Q[1](x,y,theta1,theta2), Q[2]=...
> map(alias,["]):
\end{verbatim}
\begin{equation}
{{Q}_{1}}={{Q}_{1}}(\,{x}, {y}, { \theta 1}, { \theta 2}
\,), {{Q}_{2}}={{Q}_{2}} ( \! \,{ \theta 1}, { \theta 2}, Q_{1}) ,
{{f}_{1}}={{f}_{1}}(\,{x}, {y}, { \theta 1}, { \theta 2}\,)
, {{f}_{2}}={{f}_{2}}(\,{x}, {y}\,)
\end{equation}
The {\bf `type/grassman`} subroutine recognizes \NC\ objects as in:
\begin{verbatim}
> [t1, t2, Q[1], Q[2], f[1], f[2]]:
> map(type,",grassman);
\end{verbatim}
\begin{equation}
[\,{\rm true}, {\rm true}, {\rm true}, {\rm true}, {\rm false},
{\rm false}\,]
\end{equation}
The {\bf parity} command can be used to determine the {\it parity} of
composite expressions:
\begin{verbatim}
> [Q[1]+Q[2],  f[1]+f[2],  Q[1]+F[2]]:
> map(parity,");
\end{verbatim}
\begin{equation}
[\,1, 0, {\it undefined}\,]
\end{equation}
Consider the \C\ $\*$ product of a parity $=1$ object times a parity $=0$
one:
\begin{verbatim}
> Q[1] &* (f[1] + f[2]);
\end{verbatim}
\begin{equation}
{{Q}_{1}}\,{\rm \&*}\, \left( \! \,{{f}_{1}} + {{f}_{2}}\, \!
 \right)
\end{equation}
The {\bf msort} command sorts this product taking into account its
commutativity:
\begin{verbatim}
> msort(");
\end{verbatim}
\begin{equation}
 \left( \! \,{{f}_{1}} + {{f}_{2}}\, \!  \right) \,{\rm \&*}\,{{Q
}_{1}}
\end{equation}
and only objects {\it having grassman-type variables} are kept inside
a $\*$ product:
\begin{verbatim}
> mexpand("");
\end{verbatim}
\begin{equation}
 \left( \! \,{{f}_{1}}\,{\rm \&*}\,{{Q}_{1}}\, \!  \right)  + {{f
}_{2}}\,{{Q}_{1}}
\end{equation}
Consider now a nested $\*$ product:
\begin{verbatim}
> ((a*Q[1] &* (b*Q[2])) &* (Q[2]+f[1]+f[2])) &* f[2];
\end{verbatim}
\begin{equation}
 \left( \! \, \left( \! \,{a}\,{{Q}_{1}}\,{\rm \&*}\,{b}\,{{Q}_{2
}}\, \!  \right) \,{\rm \&*}\, \left( \! \,{{Q}_{2}} + {{f}_{1}}
 + {{f}_{2}}\, \!  \right) \, \!  \right) \,{\rm \&*}\,{{f}_{2}}
\end{equation}
The {\bf midentity} command transform it into an ``un-nested" $\*$
product:
\begin{verbatim}
> midentity(");
\end{verbatim}
\begin{equation}
{a}\,{b}\,{{f}_{2}}\,{\rm \&*} \left( \! \,{{Q}_{1}}, {{Q}_{2}},
{{Q}_{2}} + {{f}_{1}} + {{f}_{2}}\, \!  \right)
\end{equation}
The {\bf mexpand} command automatically cancels products
such as $Q_2\  \* Q_2$:
\begin{verbatim}
> mexpand("");
\end{verbatim}
\begin{equation}
{a}\,{b}\,{{f}_{2}}\, \left( \! \,{\rm \&*} \left( \! \,{{f}_{1}}
, {{Q}_{1}}, {{Q}_{2}}\, \!  \right)  + {{f}_{2}}\, \left( \! \,{
{Q}_{1}}\,{\rm \&*}\,{{Q}_{2}}\, \!  \right) \, \!  \right)
\end{equation}
\subsubsection*{Differentiation examples}
To start with, let's consider the mixed derivatives of the
grassman function $Q_1$, w.r.t the grassman variables $\theta_1$ and
$\theta_2$, and check the anticommutativity of these derivatives:
\begin{verbatim}
> q[t1,t2] := gdiff(Q[1],t1,t2);
\end{verbatim}
\begin{equation}
{{q}_{{ \theta 1}, { \theta 2}}} := {\frac {{ \partial}^{2}}{{
\partial}{ \theta 2}\,{ \partial}{ \theta 1}}}\,{{Q}_{1}}
\end{equation}
\begin{verbatim}
> q[t2,t1] := gdiff(Q[1],t2,t1);
\end{verbatim}
\begin{equation}
{{q}_{{ \theta 2}, { \theta 1}}} := -{\frac {{ \partial}^{2}}{{
\partial}{ \theta 2}\,{ \partial}{ \theta 1}}}\,{{Q}_{1}}
\end{equation}
The conversion between the {\bf D} and the {\bf gdiff} formats is
accomplished via:
\begin{verbatim}
> useD(q[t1,t2]) = usegdiff(useD(q[t1,t2]));
\end{verbatim}
\begin{equation}
{{D}_{3, 4}} \left( \! \,{{Q}_{1}}\, \!  \right) (\,{x}, {y}, {
\theta 1}, { \theta 2}\,)
= {\frac {{ \partial}^{2}}{{ \partial}{ \theta 2}\,{ \partial}{
\theta 1}}}\,{{Q}_{1}}
\end{equation}
Derivatives w.r.t \C\ variables are displayed separately:
\begin{verbatim}
> gdiff(Q[1],t2,x,t1,y);
\end{verbatim}
\begin{equation}
-{\frac {{ \partial}^{2}}{{ \partial}{ \theta 2}\,{ \partial}{
\theta 1}}}\, \left( \! \,{\frac {{ \partial}^{2}}{{ \partial}{y}
\,{ \partial}{x}}}\,{{Q}_{1}}\, \!  \right)
\end{equation}
Finally, consider the second order derivatives of the $\*$ product of
$Q_1$ with the composite function
$Q_{2}(\theta 1,\theta 2,Q_{1})$:
\begin{verbatim}
> q[0] := Q[1] &* Q[2]:
> gdiff(",t1);
\end{verbatim}
\begin{eqnarray}
\lefteqn{ \left( \! \, \left( \! \,{\frac {{ \partial}}{{
\partial}{ \theta 1}}}\,{{Q}_{1}}\, \!  \right) \,{\rm \&*}\,{{Q}
_{2}}\, \!  \right)   } \\
 & &
-  \left( {\vrule
height0.79em width0em depth0.79em} \right. \! \!
\left( \! \,{{D}_{1}} \left( \! \,{{Q}_{2}}\, \!  \right)
\left( \! \,{ \theta 1}, { \theta 2}, {{Q}_{1}}\, \!  \right)
+  \left( \! \,{{D}_{3}} \left( \! \,{{Q}_{2}}\, \!  \right)
\left( \! \,{ \theta 1}, { \theta 2}, {{Q}_{1}}\, \!  \right) \,
{\rm \&*}\, \left( \! \,{\frac {{ \partial}}{{ \partial}{ \theta
1}}}\,{{Q}_{1}}\, \!  \right) \, \!  \right) \, \!  \right) \,
{\rm \&*}
{{Q}_{1}} \! \! \left. {\vrule
height0.79em width0em depth0.79em} \right)
\nonumber
\end{eqnarray}
\begin{verbatim}
> q[t1,t1] := mexpand(gdiff(",t1));
\end{verbatim}
\begin{equation}
{{q}_{{ \theta 1}, { \theta 1}}} := 0
\end{equation}
\begin{verbatim}
> gdiff(q[0],t1,t2) + gdiff(q[0],t2,t1);
\end{verbatim}
\begin{equation}
0
\end{equation}

\section{Conclusions}

The main idea of this paper is the possible extension of the computational
domain of general-purpose symbolic systems to include ACVs. This extension
was here presented, for the Maple system, as a set of type subroutines and
commands for evaluating products and derivatives taking into account the
anti-commutative character of the symbols involved.

Concerning the compatibility of the new routines with the system as a
whole, it is good, since almost no any assignments were made to standard
routines, except for the {\bf `diff/function`} subroutine. On the other
hand, some commands of the MSL may attempt to convert {\bf D} derivatives
to the {\bf diff} format while making calculations, thus introducing
errors when high order derivatives w.r.t anti-commutative variables are
involved.

To conclude, this work can be extended by introducing symbols to represent
indexed grassman functions associated with spinor representations of
space-time, for instance, including the rules for manipulating the
$\gamma^\mu$ Dirac matrices. Such an implementation would be relevant in
the introduction of general-purpose symbolic computing systems in
theoretical physics research.

\section*{Acknowledgments}

\noindent This work was supported by the State University of Rio de
Janeiro (UERJ) and the National Research Council (CNPq) of Brazil. The
author would like to thank Ali Ayari, from the Centre de recherches
math\'ematiques, Universit\'e de Montr\'eal for useful discussions.

\label{lastpage}
\end{document}